\documentstyle[prb,aps]{revtex}
\begin{document}

\title{Longitudinal Current Dissipation in Bose-glass Superconductors}
\author{David R. Nelson\ $^1$ and Leo Radzihovsky\ $^2$}
\address{Department of Physics, Harvard University, Cambridge, MA
02138\ $^1$}
\address{Department of Physics, University of Colorado, Boulder, 
CO 80309\ $^2$}
\date{\today}
\maketitle
\begin{abstract}
A scaling theory of vortex motion in Bose glass superconductors with
currents parallel to the common direction of the magnetic field and
columnar defects is presented. Above the Bose-glass transition the
longitudinal DC resistivity $\rho_{||}(T)\sim (T-T_{BG})^{\nu' z'}$
vanishes much faster than the corresponding transverse resistivity
$\rho_{\perp}(T)\sim (T-T_{BG})^{\nu' (z'-2)}$, thus {\it reversing}
the usual anisotropy of electrical transport in the normal state of
layered superconductors.  In the presence of a current $\bf J$ at an
angle $\theta_J$ with the common field and columnar defect axis, the
electric field angle $\theta_E$ approaches $\pi/2$ as $T\rightarrow
T_{BG}^+$.  Scaling also predicts the behavior of penetration depths
for the AC currents as $T\rightarrow T_{BG}^-$, and implies a {\it jump
discontinuity} at $T_{BG}$ in the superfluid density describing
transport parallel to the columns.

\end{abstract}
\pacs{PACS: 64.60.Fr, 74.20.D}


Recently there have been many efforts to understand the nature of
vortex states and dissipation in disordered high temperature
superconductors.\cite{reviews} These efforts have led to
predictions\cite{FFH,NV} that the linear resistivity does in fact
vanish at a finite transition temperature to a glassy vortex state, in
contrast to the traditional Anderson-Kim picture which always admits
small but finite linear resistivity.  There is now
general agreement on the possibility of a true linearly dissipationless
vortex state, and the theory continues in a state of active
development.

Pinning in superconductors comes in the form of point disorder such as
oxygen vacancies and interstitials as well as correlated disorder such
as screw dislocations, twin planes and artificially introduced
columnar defects. It was originally proposed\cite{FFH} that point-like
disorder would lead to a vortex glass phase, while the theory in the
presence of columnar defects (correlated disorder) predicted an
anisotropic ``Bose glass'' phase,\cite{NV} so called because of an
analogy with the theory of bosons in superfluids on disordered
substrates.\cite{BG} Although the general phenomena of divergent
pinning barriers for vanishing currents underpins both the vortex
glass and the Bose glass theories, the two theories can
and have been qualitatively distinguished experimentally via their
predictions for the transverse field $H_\perp$ response\cite{NV},
i.e. tilting of the applied magnetic field. While the vortex glass
hypothesis predicts isotropic response functions that are nonsingular
as $H_\perp\rightarrow 0$, Bose glass theory predicts a transverse
Meissner effect, with a divergent tilt modulus $c_{44}$ and a
cusp-like phase boundary in the $T-H_\perp$ phase
diagram\cite{NV}. More recently the very existence of the
three-dimensional vortex glass phase has been called into question, by
computer simulations with finite screening,\cite{Young} and by
experiments that find a first-order transition in the detwinned
samples,\cite{Safar} removing the natural source of correlated
disorder. Moreover, experiments that use electron irradiation to
inject point centers in sufficient quantities to destroy this first
order transition find no evidence for a sharp phase transition with
universal exponents.\cite{Fendrich} Nevertheless, establishing whether
the correlated or point disorder controls the low temperature
physics in a given sample remains an open and important
question.\cite{comment}

Most experimental work on glassy vortex states has focussed on current
transport perpendicular to the magnetic field and in the case of Bose
glass, perpendicular to the columnar defect axis. An exception is the
work by Seow et. al.\cite{Seow}, which measures electrical transport
parallel to the field direction in Bi$_2$Sr$_2$CaCu$_2$O$_8$ single
crystals, irradiated with heavy ions to produce columnar defects, also
along the field direction. In this note we analyze the dissipation in
the Bose glass superconductor, generalizing the scaling theory to
include both longitudinal and transverse currents. Thus, measurements
in the simultaneous presence of both longitudinal and transverse
currents also provide a clear {\it qualitative} distinction between the
vortex glass and Bose glass scenarios. When the theory is applied to AC
conductivity below $T_{BG}$, we find finite penetration depths
parallel and perpendicular to the columns. Scaling predicts a
{\it discontinuous jump} to zero of the condensate superfluid density
describing transport parallel to the columns as $T\rightarrow T_{BG}^-$.

We assume point disorder can be neglected at high temperatures and
consider a current $J$ at a finite angle $\theta_J$ with the $||$-axis
defined by the columnar defects and the magnetic field $B$. Following
the usual assumption of a scaling theory that near a continuous
transition the diverging correlation length is the only important
length scale we determines the temperature dependence and the relation
between all the physical quantities and in particular the IV
characteristics. Near a Bose glass transition dominated by columnar
defects there are two divergent correlation lengths
$l_\perp\sim|T-T_{BG}|^{-\nu'}$ and
$l_{||}\sim|T-T_{BG}|^{-\nu_{||}'}$ and a correlation time $\tau\sim
l_\perp^{z'}\sim|T-T_{BG}|^{-z'\nu'}$, where within the Bose glass
phase $l_\perp$ and $l_{||}$ measure the corresponding localization
lengths of the vortex lines. Following Ref.\cite{NV} dimensional
analysis allows us to relate physical quantities to these correlations
lengths. In three dimensions the free energy density scales as $f\sim
1/(l_\perp^2 l_{||})$. Analogously gauge invariance of the
Ginzburg-Landau theory implies that the fluctuating vector potential
scales according to
\begin{mathletters}
\begin{eqnarray}
{A}_\perp&\sim& {1\over l_\perp(T)}\;,\\
A_{||}&\sim& {1\over l_{||}(T)}\;.
\end{eqnarray}%
\label{A}
\end{mathletters}%
The definitions of the current ${\bf{J}}={\partial f/\partial{\bf{A}}}$
and the electric field ${\bf{E}}=-{\partial{\bf{A}}/\partial t}$ and
Eqs.\ref{A} allow us also to express $\bf{J}$ and $\bf{E}$ in terms
of correlations lengths and time,
\begin{mathletters}
\begin{eqnarray}
J_\perp&\sim& {1\over l_\perp l_{||}}\;,\\
J_{||}&\sim& {1\over l_\perp^2}\;,
\end{eqnarray}
\label{J}%
\end{mathletters}%
and
\begin{mathletters}
\begin{eqnarray}
E_\perp&\sim& {1\over l_\perp^{1+z'}}\;,\\
E_{||}&\sim& {1\over l_{||} l_\perp^{z'}}\;,
\end{eqnarray}
\label{E}%
\end{mathletters}%
where the relation between the correlation time and length
$\tau\sim l_\perp^{z'}$ was used. Given the above dependences of $\bf E$ and
$\bf J$ we can construct a relation between them, the IV curve, by
equating the appropriate dimensionless quantities. Upon first
considering separately currents parallel and perpendicular to the
field direction, we have
\begin{mathletters}
\begin{eqnarray}
E_\perp l_\perp^{1+z'}&\sim& F_\pm^\perp(l_\perp l_\parallel
J_\perp\phi_o/c T)\;,\\ 
E_{||} l_\parallel l_\perp^{z'}&\sim& F_\pm^\parallel(l_\perp^2
J_\parallel\phi_o/c T)\;,
\end{eqnarray}
\label{IV}%
\end{mathletters}%
where $\phi_0=2\pi\hbar c/2e$ is the flux quantum and we have set $k_B=1$.
The dimensionless arguments of the scaling functions $F_\pm^\perp$ and
$F_\pm^\parallel$ are the ratios of the work done by the corresponding
current to depin the vortex line from the columnar defect to the
thermal energy. The difference in the arguments can be understood
microscopically.  For a transverse current $J_\perp$ dissipation
arises due the vortex line depinning which proceeds via a
``tongue'' of a typical area $l_\perp\times l_\parallel$ lying in the
$z-r_\perp$-plane. In contrast for a longitudinal current
$J_\parallel$ the dissipation is due to depinning of vortex helices
whose projections span a typical area $l_\perp^2$ lying in the
${\bf r}_\perp$-plane. 

The scaling functions above $F_+$ and below $F_-$ the transition are
very different. For $T>T_{BG}$ we expect linear resistivities
$E_\perp=\rho_\perp J_\perp$ and $E_\parallel=\rho_\parallel
J_\parallel$ characteristic of a normal metal. It follows that the
positive branches of these scaling functions must vanish linearly,
$F_+(x)\sim x$, i.e. 
\begin{mathletters}
\begin{eqnarray}
\rho_\perp&\sim& {l_\parallel/l_\perp^{z'}}\;,\\
\rho_\parallel&\sim& {1/(l_\perp^{z'-2}l_\parallel)}\;.
\end{eqnarray}
\label{rho}%
\end{mathletters}%
There is excellent theoretical\cite{BG,NV} and
numerical\cite{Krauth,Wallin} evidence that vortices in the liquid
phase (i.e. the ``superfluid''state of the bosons) ``diffuse'' as they
wander along the average field direction. This implies an important
relation between the localization lengths near Bose glass\cite{NV}
transition $l_\parallel\approx (T B^2/c_{11}\phi_o^2) l_\perp^2$,
where $c_{11}\approx B^2/8\pi$ ($B>>H_{c1}$) is the bulk modulus of
the vortex liquid. Using these relations together with the temperature
dependence of $l_\perp$ in Eqs.\ref{rho} we find,
\begin{mathletters}
\begin{eqnarray}
\rho_\perp(T)&\sim& |T-T_{BG}|^{\nu'(z'-2)}\;,\\
\rho_\parallel(T)&\sim&|T-T_{BG}|^{\nu'z'}\;.\
\end{eqnarray}
\label{rhoT}%
\end{mathletters}%
Close to the transition, $\rho_{\parallel}<<\rho_\perp$, which is {\it
opposite} to the usual normal state resistivity anisotropy in layered
superconductors. 

Consider a current $\bf J$ at an angle $\theta_J$ with
the B field and columnar defect axis. There is now a matrix relating
$\bf E$ to $\bf J$,
\begin{equation}
\left[ \begin{array}{c}
E_\perp \\
E_\parallel \end{array} \right]
\approx
\left[ \begin{array}{cc}
\rho_\perp & 0 \\
0 & \rho_\parallel \end{array} \right]
\left[ \begin{array}{c}
J_\perp \\
J_\parallel \end{array} \right]\;,
\end{equation}
where the off-diagonal elements are zero if we neglect the very small
and poorly understood Hall effect. The electric field $\bf E$ (in this
single parameter scaling theory) will be at a temperature-dependent
angle $\theta_E(T)=\tan^{-1}(E_\perp/E_\parallel)$, given by
\begin{equation}
\tan(\theta_E)\propto\tan(\theta_J)/(T-T_{BG})^{2\nu'}\;,
\label{thetaE}
\end{equation}
where $\tan(\theta_J)=J_\perp/J_\parallel$.  Equation \ref{thetaE}
predicts that near $T_{BG}$ the angle $\theta_E$ for Bose glass
superconductor has a universal temperature dependence controlled by
the Bose glass transverse localization length exponent $\nu'$,
estimated to be $\nu'\approx 1$.\cite{Krauth,Wallin} Besides providing
a direct measurement of $\nu'$, Eq.\ref{thetaE} predicts the electric
field direction $\theta_E(T\rightarrow T^+_{BG})\approx \pi/2 -
(T-T_{BG})^{2\nu'}\cot(\theta_J)\rightarrow\pi/2$, for any current
direction $\theta_J\neq 0$, independent of microscopic details such as
the intrinsic resistivity anisotropy of the normal state.  Because
vortex glass dissipation is isotropic (aside from the intrinsic
material anisotropy) the corresponding expression for vortex glass
predicts a $\theta_E$ that is asymptotically temperature independent
as $T\rightarrow T_{VG}$ and depends continuously on the direction of
the current $\theta_J$.

The significantly faster vanishing of longitudinal resistivity,
predicted by Eq.\ref{rhoT} as $T\rightarrow T_{BG}$ has already been
observed in recent experiments by Seow, et al.\cite{Seow}, which finds
$\nu' z' = 8.5 \pm 1.6$, consistent with other estimates of
$\nu'=1$\cite{Krauth,Wallin} and $z'= 6.0\pm 0.5$ \cite{Wallin}.
However, as is evident from Eq.\ref{thetaE}, an additional check on 
the Bose glass theory can be made by 
testing to see if $\lim_{T\rightarrow T_{BG}^+}\theta_E(T)=\pi/2$ for
any $\theta_J\neq 0$. Equivalently, the measurement of the vanishing
ratio $\rho_\parallel(T)/\rho_\perp(T)$ as $T\rightarrow T_{BG}$ allows
a direct determination of $\nu'$.

The scaling Eq.\ref{IV} predicts nonlinear IV characteristics {\it at}
the Bose glass transition, $T=T_{BG}$.\cite{FFH,NV} The requirement
that there is a well defined IV characteristics demands that the
divergent correlation lengths cancel on both sides of these equations,
which can only be satisfied by a specific power-law behavior of 
$F^\perp(x)$ and $F^\parallel(x)$ as $x\rightarrow\infty$, leading to 
\begin{mathletters}
\begin{eqnarray}
E_\perp(J_\perp)&\sim& J_\perp^{(1+z')/3}\;,\\
E_\parallel(J_\parallel)&\sim& J_\parallel^{(2+z')/2}\;.
\end{eqnarray}
\label{IVTbg}%
\end{mathletters}%
The longitudinal dissipation is thus weaker and more nonlinear
than the transverse one.

Below the Bose glass transition the dissipation is highly nonlinear
and is characterized by potential barriers that diverge in the limit
of vanishing current,
\begin{mathletters}
\label{IVbelowTbg}
\begin{eqnarray}
E_\perp(J_\perp)&\sim& e^{-(J_\perp^o/J_\perp)^{\mu_\perp}}\;,\\
E_\parallel(J_\parallel)&\sim&
e^{-(J_\parallel^o/J_\parallel)^{\mu_\parallel}}\;,
\end{eqnarray}%
\end{mathletters}%
where $\mu_\perp\rightarrow 1/3$ as $J_\perp\rightarrow 0$ in bulk
samples\cite{NV} and the calculation in Sec.II-E of Ref.\cite{NV}
suggests that $\mu_\parallel=1$.

The scaling theory can be further generalized to a finite
frequency $\omega$ by an addition to the scaling functions in
Eqs.\ref{IV} of another dimensionless variable $\omega l_\perp^{z'}$.
At finite frequency there is linear dissipation at all finite
temperatures characterized by linear conductivities\cite{FFH,Dorsey,Yeh} 
\begin{mathletters}
\begin{eqnarray}
\sigma_\perp(\omega,T)&\sim& l_\perp^{z'-2} f_\pm^\perp(\omega
l_\perp^{z'})\;,\\ 
\sigma_\parallel(\omega,T)&\sim& l_\perp^{z'} 
f_\pm^\parallel(\omega l_\perp^{z'})\;. 
\end{eqnarray}
\label{sigma}%
\end{mathletters}%
Requiring that the conductivities are finite {\em at} the Bose glass
transition, the scaling theory together with the Kramers-Kronig
relation lead to
\begin{mathletters}
\begin{eqnarray}
\sigma_\perp(\omega,T_{BG})&\sim&\left({1\over
-i\omega}\right)^{1-2/z'}\;,\\
\sigma_\parallel(\omega,T_{BG})&\sim&{1\over
-i\omega}\;,
\end{eqnarray}
\label{sigmaii}%
\end{mathletters}%
predicting a universal phase lag between current and voltage, which in
the case of $\sigma_\parallel$ is $\pi/2$, independent of critical
exponents. For $T<T_{BG}$, $\sigma_{\perp,\parallel}\sim
n_s^{\perp,\parallel}/(-i\omega)$, implying scaling for the superfluid
number densities describing charge transport by Cooper pairs
perpendicular and parallel to the columns,
\begin{mathletters}
\begin{eqnarray}
n_s^\perp&\sim&1/l_\parallel\sim1/l_\perp^2\;,\\
n_s^\parallel&\sim&l_\parallel/l_\perp^2=\mbox{constant}\;,
\end{eqnarray}
\label{rhos}%
\end{mathletters}%
as $T\rightarrow T_{BG}^-$, consistent with the corresponding
Josephson relations for superfluid densities in an anisotropic
superconductor. 

More precisely, for $n_s^\parallel$ we expect the relationship
\begin{equation}
\lim_{T\rightarrow T_{BG}^-} n_s^\parallel(T)=\lim_{T\rightarrow T_{BG}^-}{m
T\over\hbar^2}{l_\parallel(T)\over l_\perp^2(T)}=\mbox{constant}\;,
\label{ns}
\end{equation}
where we take $m$ to be the mass of a Cooper pair, and $l_\parallel(T)$ and
$l_\perp(T)$ are defined in the usual way in terms of the decay of the
transverse BCS order parameter correlation
function\cite{Hohenberg}. We have assumed in the spirit of scaling
that $l_\parallel(T)$ and $l_\perp(T)$ are the only diverging length
scales near $T_{BG}$. The lengths $l_\parallel(T)$ and $l_\perp(T)$
must then diverge as $T\rightarrow T_{BG}^-$ in the same way as the
corresponding correlation lengths above $T_{BG}$. Since
$\lim_{T\rightarrow T_{BG}^+}l_\parallel(T)/l_\perp^2(T)\approx T_{BG}
B^2/c_{11}\phi_0^2=\mbox{constant}$, required by
finiteness of the boson compressibility $c_{11}$ (vortex line
compression modulus) at $T_{BG}$ \cite{NV,BG}, we are led to
Eq.\ref{ns}. In the likely event that for short
range interaction both superfluid densities vanish in the vortex
liquid state for $T>T_{BG}$, our analysis therefore implies a striking
result: In contrast to $n_s^\perp$ which vanishes smoothly as
$T\rightarrow T_{BG}^-$ (similar to a conventional superconductor),
$n_s^\parallel$ has a {\em jump discontinuity} at $T=T_{BG}$
analogous to a stiffness of a system at a Kosterlitz-Thouless
transition. The $n_s^\parallel$ jump discontinuity is consistent
with Eq.\ref{sigmaii}b, predicting that $\sigma_\parallel$'s $\omega$
dependences at and below $T_{BG}$ are identical. 

%
%

Using above results for the AC conductivities together with Maxwell's
equations, we find the effective penetration lengths $\lambda_{\mbox{
eff}}\sim1/\sqrt{\omega |\sigma(\omega)|}$\cite{FFH} for the AC currents
$J_\perp$ and $J_\parallel$ (for $\omega\rightarrow 0$) to be
${\tilde\lambda}_\perp\sim \sqrt{l}_\parallel\sim l_\perp$ and
${\tilde\lambda}_\parallel\sim l_\perp^2/l_\parallel=\mbox{constant}$,
respectively. While ${\tilde\lambda}_\perp$ diverges as $T\rightarrow
T_{BG}^-$, ${\tilde\lambda}_\parallel$ remains finite at the
transition and discontinuously jumps to infinity for $T>T_{BG}$.

The scaling theory for longitudinal currents can be further
generalized to include the response to the transverse magnetic field
$H_\perp$, previously analyzed for $J_\perp$ in Ref.\cite{NV}. For
simplicity assuming purely longitudinal current, $E_\parallel$ from
Eq.\ref{IV} becomes
\begin{equation}
E_{||} l_\parallel l_\perp^{z'}\sim F_\pm^\parallel(l_\perp^2
J_\parallel\phi_o/c T,\; H_\perp l_\perp l_\parallel/\phi_0)\;.
\label{IVHperp}
\end{equation}
which by arguments similar to above predicts a cusp-like phase
boundary in the T-H$_\perp$-plane between the Bose glass where
$\rho_\parallel(H_\perp<H_\perp^c(T))=0$ and the vortex liquid phase with
$\rho_\parallel(H_\perp>H_\perp^c(T))>0$. This boundary, given by
\begin{equation}
H_\perp^c(T)\sim\pm|T-T_{BG}|^{3\nu'}\;,
\end{equation}
is consistently identical to the phase boundary obtained based on the
criterion of the vanishing of the transverse resistivity
$\rho_\perp(T)$, as must be the case if there is a single transition
to the Bose glass phase.

Equation \ref{IVHperp} can also be used to predict how
$\rho_\parallel(H_\perp)$ vanishes as $H_\perp\rightarrow 0$, 
with $T=T_{BG}$,
\begin{equation}
\rho_\parallel(T = T_{BG}, H_\perp)\sim |H_\perp|^{z'/3}\;.
\end{equation}
This result is to be contrasted with the more slowly
vanishing transverse linear resistivity 
$\rho_\perp(T = T_{BG}, H_\perp)\sim |H_\perp|^{(z'-2)/3}$ found in
Ref.\cite{NV}.

We are grateful to the authors of Ref.\cite{Seow} for a preprint of
their work prior to publication.  DRN acknowledges financial support
by National Science Foundation, in part by the MRSEC program through
Grant DMR-9400396 and through Grant DMR-9417047. LR was supported by
the National Science Foundation CAREER award, through Grant
DMR-9625111.

\end{document}